# The Mediterranean deep-sea fauna: historical evolution, bathymetric variations and geographical changes

Christian C. Emig[1]

Patrick Geistdoerfer[2]

**Abstract:** The deep-water fauna of the Mediterranean is characterized by an absence of distinctive characteristics and by a relative impoverishment. Both are a result of events after the Messinian salinity crisis (Late Miocene). The three main classes of phenomena involved in producing or recording these effects are analysed and discussed:
  - Historical: Sequential faunal changes during the Pliocene and thereafter in particular those during the Quaternary glaciations and still in progress.
  - Bathymetric: Changes in the vertical aspects of the Bathyal and Abyssal zones that took place under peculiar conditions, *i.e.* homothermy, a relative oligotrophy, the barrier of the Gibraltar sill, and water mass movement. The deeper the habitat of a species in the Mediterranean, the more extensive is its distribution elsewhere.
  - Geographical: There are strong affinities and relationships between Mediterranean and Atlantic faunas. Endemic species remain a biogeographical problem. Species always become smaller in size eastward where they occupy a progressively deeper habitat.
   Thus, the existing deep Mediterranean Sea appears to be younger than any other deep-sea constituent of the World Ocean.

**Key Words:** Mediterranean Sea; deep-sea fauna; bathyal; abyssal; changes; glaciations

**Résumé :** ***Faune profonde en Mer Méditerranée : les échanges historiques, géographiques et bathymétriques.-*** Le benthos profond méditerranéen est caractérisé par une absence d'originalité et une pauvreté dont les raisons sont à rechercher dans l'histoire récente de la faune. Trois types principaux d'échanges ont été distingués :
  - les échanges historiques à travers les changements de faunes depuis le Pliocène et durant les glaciations du Quaternaire ;
  - les échanges bathymétriques au sein des étages Bathyal et Abyssal soumis à des conditions très particulières (homothermie, relative oligotrophie, barrière du seuil de Gibraltar, circulation des masses d'eaux) ; plus profonde est l'extension des espèces en Mer Méditerranée et plus large est leur distribution hors Méditerranée ;
  - les échanges géographiques avec des affinités étroites entre Mer Méditerranée et Océan Atlantique. Le cas des espèces endémiques reste un problème biogéographique. Les espèces ont toujours une distribution plus profonde en allant vers l'Est et leur taille devient plus petite.
   Ainsi, la Mer Méditerranée profonde actuelle apparaît comme une mer beaucoup plus jeune qu'aucune autre partie de l'Océan mondial profond.

**Mots-Clefs :** Mer Méditerranée ; faune profonde ; bathyal ; abyssal ; échanges ; glaciations

## Introduction

The shelf-break of the continental shelf is a barrier separating the neritic domain from the oceanic deep-sea domain, and is as important a limit as the coast line. It is defined by geological, physico-chemical and biological characteristics (EMIG, 1997).

The deep-sea domain is divided in three zone (Table 1). Their areal extent in the Mediterranean Sea differs from those in the Ocean. The Mediterranean deep-sea fauna remains poorly known probably because of the absence of peculiarities and a relative impoverishment, which may be explained by the recent history of the fauna. Three discrete variables are involved:

(i) historical evolution that began in Late Miocene times and was especially effective during the climatic fluctuations of the Quaternary with its succession of glaciations (Fig. 1);

(ii) vertical changes resulting from the very particular conditions of the deep-sea environment;

---
[1]   CNRS UMR 6540, Centre d'Océanologie, Rue de la Batterie-des-Lions, 13007 Marseille (France)
      e-mail: Christian.Emig@com.univ-mrs.fr
[2]   CNRS UMR 7093, Laboratoire d'Océanographie de Villefranche; Laboratoire d'Ichtyologie, Muséum National d'Histoire Naturelle, 43 Rue Cuvier, 75231 Paris cedex 05 (France)
      e-mail: Geist@mnhn.fr

(iii) geographic influences involving changes in the relationships of the western and eastern Mediterranean basins, and also between the Mediterranean and the adjacent portion of the Atlantic Ocean.

|  | Neritic Domain | Deep-Sea Oceanic Domain | | |
|---|---|---|---|---|
|  | Continental Shelf | Bathyal Zone | Abyssal Zone | Hadal Zone |
| **Limits in m** | | | | |
| World Ocean | 0 m to Shelf-break | Shelf-break to 3,000 m | 3,000 to 6,000 m | More than 6,000 m |
| **Mediterranean Sea** | 0 to 100-110 m | 100-110 to 3,000 m | 3,000 to 5,093 m | - |
| **Extension in % of the entire area** | | | | |
| World Ocean | 7 % | 15 % | 77 % | 1 % |
| **Mediterranean Sea** | 15 % | 72 % | 13 % | - |

**Table 1:** Comparison of the ranges in depth of the deep-sea zones of the World Ocean and the Mediterranean Sea. The Mediterranean comprises 0.82 % of the surface of the World Ocean and 0.35 % of its volume. The edge of the continental shelf is the boundary between the neritic (inshore) domain and the deep-sea oceanic (offshore) domain. Its depth varies with the ocean or sea.

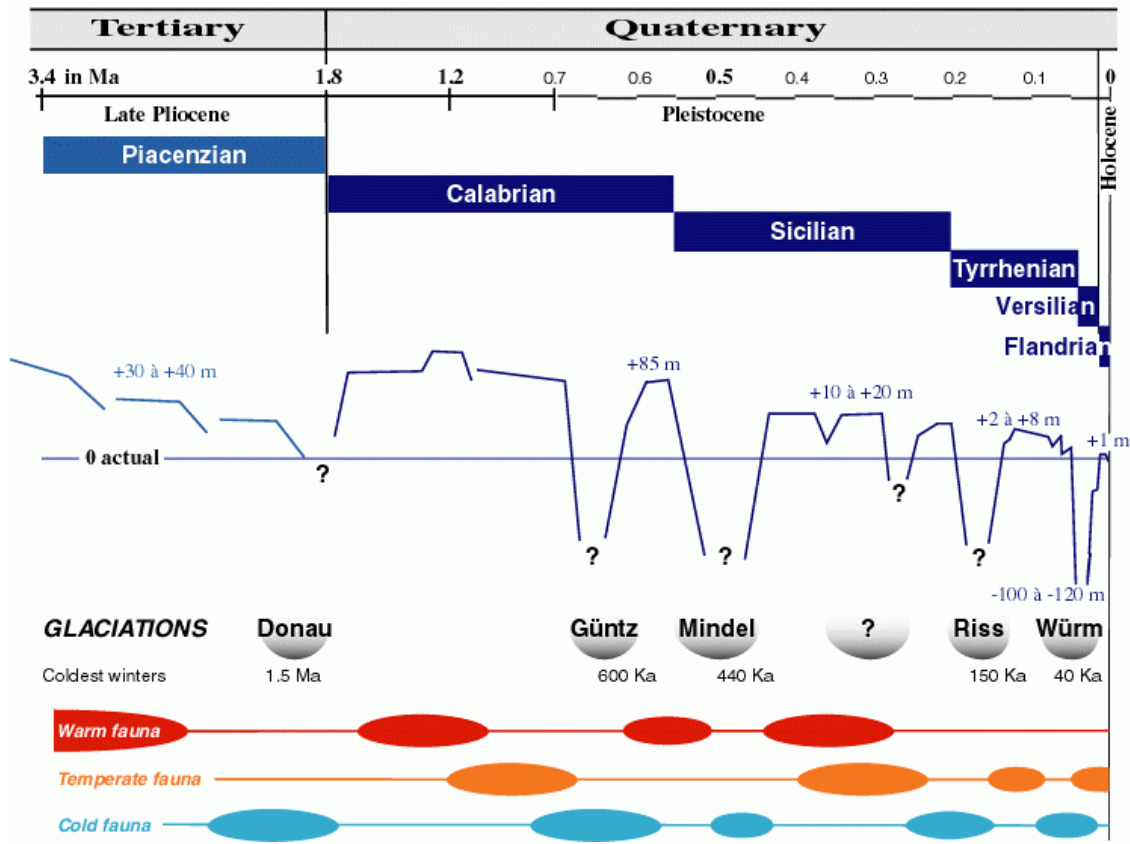

**Figure 1:** Main faunal successions in the Mediterranean Sea during the several glaciations after the Pliocene (modified, from LAUBIER & EMIG, 1993). The true number of glaciations during the Pleistocene remains to be defined, there are probably one or two more between the Mindel and Riss glaciations.

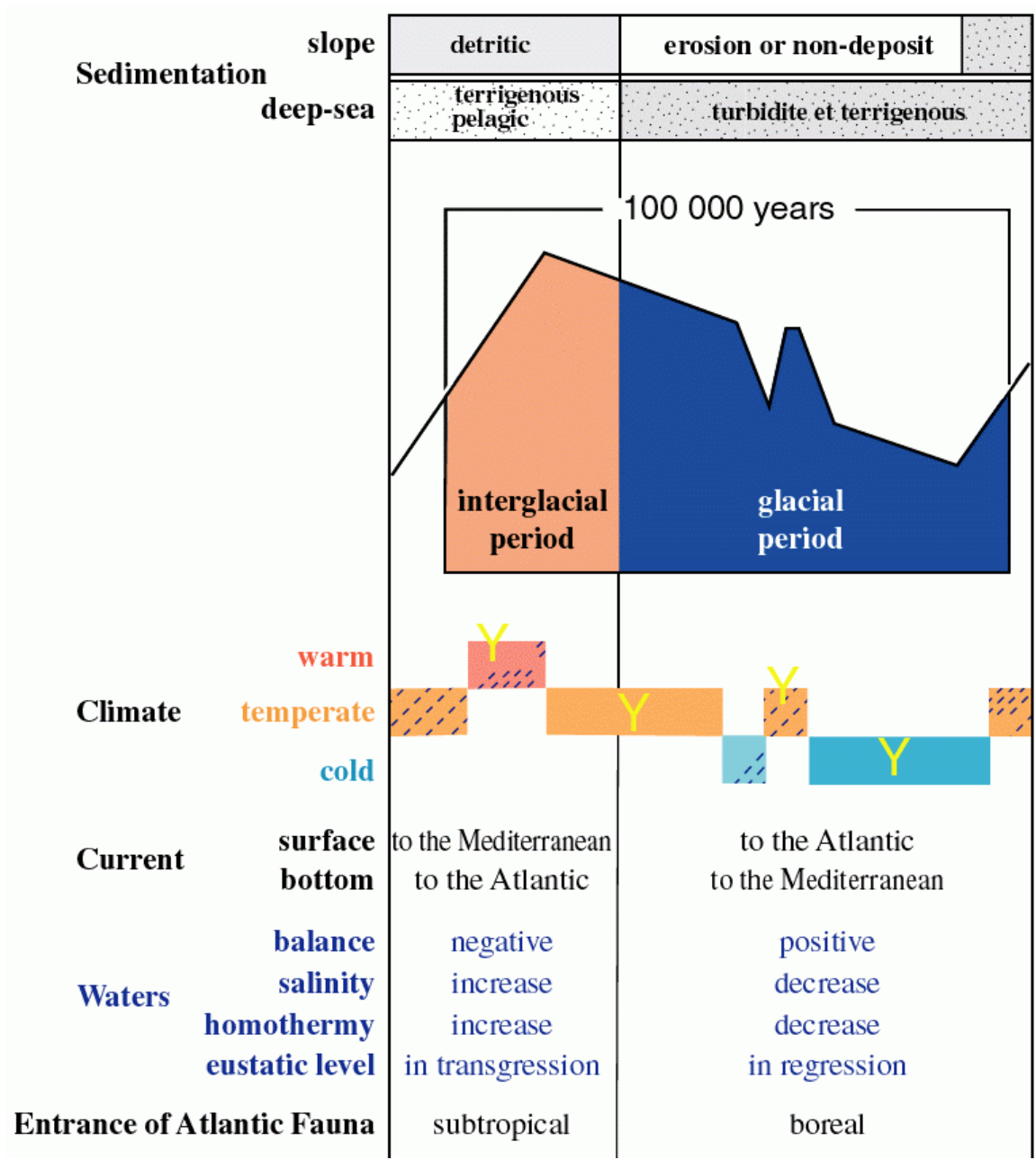

**Figure 2:** Summary of the principal characteristics of a Quaternary glaciation cycle (glacial and interglacial period) in the Mediterranean Sea.

All of these factors are directly influenced by the ecological characteristics of the Mediterranean deep-sea, predominantly by its homothermy, varying eastwards from 13 to 13.5°C in the Western basin and from 13.5 to 15.5°C in the Eastern one, by its high salinity, from about 38 to 39.5 psu∗, by the stratification of the water layers and their thermohaline (barotropic) motions; and by the system of currents above the Gibraltar and Siculo-Tunisian sills (Fig. 2).

   \* According to international conventions (UNESCO, 1985), the values of salinity have no unit and are expressed in psu (= practical salinity units).

## Historical evolution

An understanding of the evolution of the Mediterranean deep-sea fauna requires consideration of the recent geological history of the sea itself, in particular after the definitive closure of its connection with the Indian Ocean during the middle Burdigalian (Early Miocene) (STEINIGER and RÖGL, 1984). The existing contours of the Mediterranean Sea were attained only at the end of the Tertiary.

The most important crisis in the recent history of the Mediterranean Sea occurred during the Messinian (Late Miocene). Called the "Messinian salinity crisis" (Hsü *et alii*, 1978) it resulted in a massive decrease in the representatives of most phyla, and especially those in the deeper waters. Consequently, most of the paleo-Mediterranean species, especially the early Neogene tropical fauna (*e.g.* Eocene-Oligocene fish, brachiopods, *etc*.) disappeared. Nevertheless, several genera and species of this period are present today in the Mediterranean fauna. The most reasonable hypothesis to explain this anomaly, but one not widely accepted (see Rouchy, 1986; Clauzon *et alii*, 1996; Blanc, 2000; Taviani, 2002) postulates that, during this crisis, there were bathyal areas (over 1,000-1,500 m depth) in which a part of the deep-sea fauna remained or survived (Ben Moussa *et alii*, 1988; Barrier *et alii*, 1989; Di Geronimo 1990; Laubier and Emig, 1993), particularly in peripheral basins like southeastern Spain (Riding *et alii*, 1999) and southern Italy (Barrier *et alii*, 1989). Refuting the theory of a desiccated deep basin, Busson (1984), Jauzein (1984), Jauzein and Hubert (1984), Busson (1990) proposed another model to explain the survival of a deep-sea fauna during the Messinian: Evaporites were deposited only intermittently in a series of oscillating basins which, during the Messinian, would have maintained an unbroken connection with the Atlantic Ocean above the Gibraltar sill and thus assured the continuous existence of abyssal plain environments. In any event, only after this "Messinian salinity crisis", did the surviving fauna began its evolution in the restored Mediterranean.

During the Plaisancian transgression (Fig. 1), the warm climate (approximately 4°C above the present) was relatively stable; warm to subtropical sea-waters predominated. This allowed the entrance through the strait of Gibraltar of an Atlantic fauna from the Ibero-Moroccan region that resulted in the first restocking of the Mediterranean Sea. Then, at about 2.1-2 Ma, the climate became temperate to cold. The change favored the entrance of a great number of species from the temperate regions of the Atlantic Ocean. An "Atlantico-Mediterranean" stock was established and became dominant in the Mediterranean fauna. Although the Plio-Pleistocene boundary remains controversial, it may be defined by the arrival of these boreal species in the Mediterranean consequent on climatic and oceanographic changes during the regression at the end of the Plaisancian. The temperature of surface waters decreased several degrees with changes in the direction of the prevailing wind during the Donau glaciation (Fig. 1).

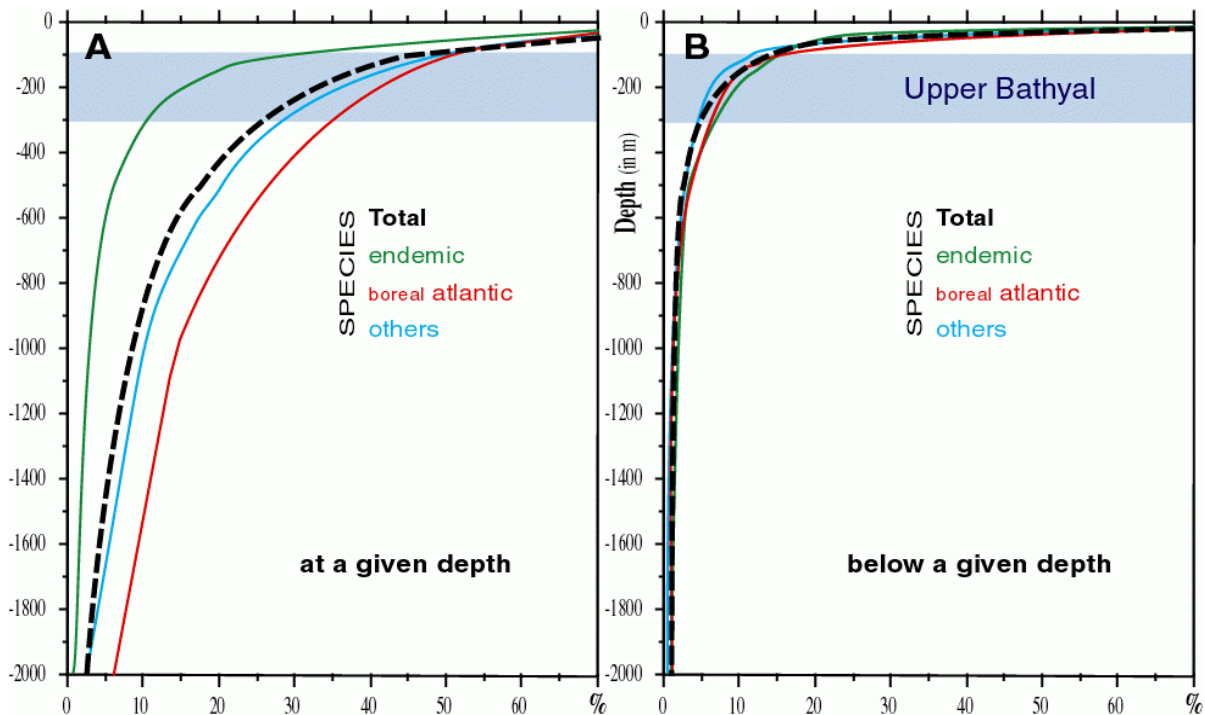

**Figure 3:** Vertical distribution of 3613 benthic invertebrate species in the Mediterranean Sea: A. percentage of the three categories at a given depth; B. percentage of species living below a given depth (data from Médifaune, FREDJ & LAUBIER, 1983).

| Age in Ka BP | -18 | -15 | -12 | -10 | -0.8 | Present |
|---|---|---|---|---|---|---|
| Sea level (- m) | 120 | 90 | 70 | 50 | 20 | 0 |
| Temperature (°C) | 8 | 9 | 10 | 8.5 | 13 | 13 |
| Salinity (psu) | 40.1 | 39 | 38.5 | 38,9 | 37 | 37.9 |
| Renewal period (years) | 133 | 124 | 91 | 79 | 97 | 75 |

**Table 2:** Paleooceanographic evolution of the western basin since the last Würm glaciation (after POUTIERS, 1987).

Thereafter, during the Quaternary (Pleistocene to Present) a succession of remarkable climatic fluctuations took place, produced by successive glacial and interglacial periods, summarized on figure 2. The bathymetry of the Mediterranean was similar to that of today, the Strait of Gibraltar was silled at about 300 m and variations in sea levels were produced by the glacio-eustatic processes of the World Ocean.

During the cold climates of the glacial periods, large biocenoses developed on the outer part of the continental shelf and along the continental slope. They consisted mainly of circalittoral and bathyal species, generally of a temperate to boreal origin. Their introduction from the Atlantic was favoured by the incoming bottom-current in the strait of Gibraltar. On the other hand, during the interglacial periods with a temperate to warm climate (representing only about 10 % of an entire cycle of glaciation: 90 to 125 Ka), the coastal current entering the Mediterranean allowed the colonisation of subtropical species from the Senegalese Province especially those in the littoral waters, while the outgoing bottom-current made the arrival of deep-sea denizens difficult (Fig. 1-2).

During the Calabrian (Donau glaciation), the Sicilian (Mindel glaciation) and the Tyrrhenian (Riss glaciation), the Pliocene stock of subtropical species disappeared and a stock of boreal species (sometimes called the "Celtic" fauna) was added in successive waves to the previous Atlantico-Mediterranean stock of temperate origin from the beginning of the Donau glaciation, that included the few surviving representatives of the paleo-Mediterranean epoch. These new arrivals took place in a temperate climate with well marked seasons. Surface water temperatures ranged between 9-10°C in winter to 19-20°C in summer. Some of these boreal species still live in the northeastern Atlantic Ocean, among them is the gastropod *Buccinum undatum* (it is still found in the Mediterranean below 1,000 m) and the bivalves *Cyprina* (*Arctica*) *islandica*, *Mya truncata*, *Panomya spengleri*, *Modiolus modiolus* and *Chlamys islandica*, and probably along with them two fishes: the pleuronectid *Platichthys flesus* and the rajid *Raia clavata*, both still netted in the Mediterranean (PÉRÈS and PICARD, 1964; RAFFI, 1986). It was at the end of the Sicilian and at the beginning of the Tyrrhenian (Riss glaciation) that the bathyal and abyssal biocenoses developed and colonized the Mediterranean.

During the Riss-Würm interglaciation (Tyrrhenian; Fig. 1), sea level was 2 to 8 m above the present level; temperate-warm species arrived, mainly from the Senegalese province, among them an important stock of molluscs no longer present.

During the last glaciation, the Würm (approximate duration from 70,000 to 18,000 years BP), sea level dropped 100 to 120 m below the present level (Fig. 1; Table 2) Living conditions for the bathyal and abyssal biocenoses were very harsh and entrance for Atlantic deep-sea species was limited. The thanatocenoses of the Celtic malacological fauna, *i.e.* the bivalve mollusc *Chlamys septemradiata* and *Cyprina* (*Arctica*) *islandica*), were contemporaneous with the Würm glaciation and are dated between 13,000 to 9,800 ±300 years BP; FROGET, 1974). Now, these two species occur between 90 and 600 m, commonly between 200 to 300 m, because they developed sublittorally when sea level was approximately 80 m lower. *Chlamys septemradiata* existed also in the Aegean Sea, but has disappeared from the Mediterranean, except in the Alboran Sea where it is commonly recorded. This species does not tolerate salinity over 36.5 psu an explanation for its limited distribution eastward. The high salinity of the Western Mediterranean has also eradicated *Cyprina* (*Arctica*) *islandica*, although it is common in fossil records.

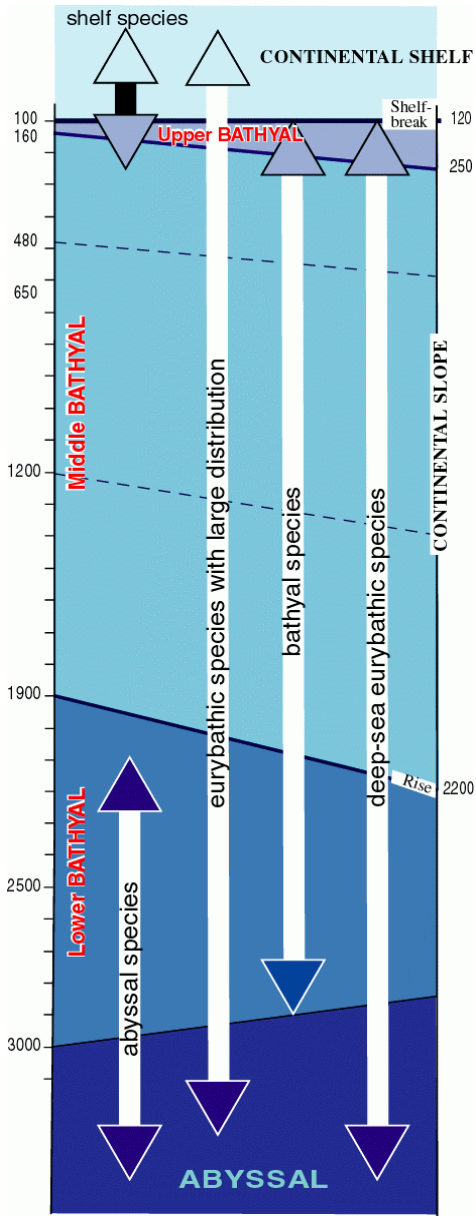

**Figure 4:** Vertical distribution of the several categories of species between the outer continental shelf and the Abyssal zone (modified, from CARPINE, 1970 ; REYSS, 1970). The range of the bathymetric limits of Bathyal zone and subzones on the continental slope and of the Abyssal zone are given (dotted: the limits of the subdivisions of the middle Bathyal zone).

In the Eastern Mediterranean, the formations of sapropels (between 2.5 to 70 cm thick, BUCKLEY and JOHNSON, 1988) took place between 25,000 and 23,000 years BP ($S_2$) and 9,000 and 6,800 years BP ($S_1$). They were deposited in association with massive flows of fresh-water from the Nile and the Black Sea that caused a stratification: that is a layer of fresh-water on highly saline cold water. At the same time, large quantities of organic and terrigenous material (partly of aeolian origin) sank to the sea floor so that all free oxygen in the stagnant bottom waters was consumed. This anoxia developed in the Bathyal zone (600-1,000 m) during an interglacial period, and at that time prevented development of a colonisation there, which in turn led to faunal extinction below this limit and thus hindered deeper colonisations. Recolonisation occurred during the glacial period (GALIL and GOREN, 1992). However, the higher temperature and salinity of the eastern basin prevented the penetration and propagation of stenothermal and stenohaline species from the western Circalittoral and Bathyal zones despite a favourable bottom-current over the Siculo-Tunisian sill.

According to ROHLING (1994), the formation of the sapropels in the eastern basin occurred under the influence of an anti-estuarine circulation due to a lessening of the previously excessive evaporation of fresh-water input (Nile) caused by an increase in the number and intensity of monsoons from the Indian Ocean that augmented the flow from the Nile. In the western basin the rise in precipitation reduced evaporation, but the absence of sapropels can be explained by the absence of layering.

According to PÉRÈS and PICARD (1964) the evolution of Mediterranean deep-water populations during the Plio-Quaternary period appears to be linked more closely to changes in the system of currents in the strait of Gibraltar than to variations in water temperatures: at least as an explanation of faunal waves derived in succession from boreal (Celtic) species and tropical to warm temperate clades. For others authors (see CARPINE, 1970), salinity and hydrodynamics, rather than temperature, governed the penetration of "warm" and "cold" faunas. Deep-water colonization is confined essentially to the Bathyal zone and the introduction and development of new elements took place mainly during the glacial periods when abiotic conditions, *i.e.* temperature, salinity, nutritional input, water currents..., favoured colonisation by "cold-temperate" species: their successful arrival depended on their characteristics and the ecological requirements in relation to available environments.

## Vertical variations

The most important factor controlling the vertical distribution of the Mediterranean deep-sea fauna is the homothermy that exists from around 200-300 m to the bottom. Its temperature increases West to East from about 13° to 15.5°C: consequently, no thermal boundaries exist in the deep-sea, apart from the homothermy itself, while in the Atlantic Ocean the temperature decreases with depth. Many species have a large bathymetric distribution in the Mediterranean not because they are eurybathic but because they are eurythermal or warm stenothermal. Thus, the distributional limits of the deep-sea faunas are governed by other factors, such as salinity, grain-size distribution, pressure, food, hydrodynamics. And this variety in environmental conditions causes the Mediterranean Bathyal fauna to be more heterogeneous than that of the Atlantic (EMIG, 1989; LAUBIER and EMIG, 1993).

The bathyal biocenoses of the Mediterranean are similar to those on the continental slope of the northeastern Atlantic, but comparatively impoverished in the number of species and in their abundance because:

- The difficulty encountered by the several species in entering the Mediterranean over the Gibraltar sill with its depth of only 300 m, and with a contrary bottom-current, both warm and strongly saline.

- The difficulty in survival owing to the homothermy and the lack of food in the deeper Mediterranean, which exhibits a relative oligotrophy strengthened by the shelf-break barrier, and in the western basin by the cyclonic current which is stronger in the Upper Bathyal zone (see EMIG 1997, with references).

The Bathyal zone is divided in several subzones. Their limits in the Mediterranean (Fig. 4) correspond to those described in the Oceanic Bathyal zone, but with modifications in their faunas. The main factors appear to be the availability of food coupled with the extra energy required to find it because of the decrease in the numbers of prey with increasing depth: mainly endobenthos and macroplancton (EMIG, 1997).

Although the Mediterranean abyssal macrobenthos is constituted by a large number of eurybathic species, there are but 20 to 30 true abyssal species (Fig. 3A). In the western basin where the depth does not exceed 3,000 m, the abyssal fauna is less abundant than in the deeper eastern basin; in the Matapan trench (-5,093 m), abyssal species are dominant. The Mediterranean macrofauna is represented by about 7,200 species (5.6 % of the world marine fauna), but this fauna is far from completely catalogued, although it is probably the best-known of all the faunas that have been and are being studied (BIANCHI and MORRI, 2000).

|  | Western Basin | Eastern Basin | | | | Black Sea |
|---|---|---|---|---|---|---|
|  |  | Adriatic | Ionian | Levantine | Aegean |  |
| Total of species in % | 92 | 54 | | | | - |
| Deep-sea species in % | 97 | 33 | | 20 | | - |
| Total of Polychaete species | 884 | 527 | 528 | 451 | 597 | 310 |
| and total in % | 85 | 51 | 51 | 44 | 57 | 30 |

**Table 3:** Impoverishment of the fauna from West to East (total known species – data compiled from various authors; for the polychaetes data from ARVANITIDIS *et alii*, 2002).

A comparison between the bathymetric distribution (Fig. 3A) of 3613 species and the distribution of those existing only below a given depth (Fig. 3B) provides an estimate of the degree of eurybathy of the fauna at any one depth (FREDJ & LAUBIER, 1983). The percentage of eurybaths in the total decreases gradually with depth, *i.e.* 25 % of them live more than 50 m down, 15 % at more than 100 m, but only 1 % below 2,000 m. Furthermore, with increments of bottom depths ranging up to 2000 m, the percentage of species that live both in littoral waters and at any given depth increases: for example, about 40 % of the species living between 0 and 50 m have a more or less extensive depth range, while below 500 m, for each deep-sea species restricted to that depth there are six that live at higher levels too. This eurybathic distribution is coincident with like distributions elsewhere, most obviously in the Atlantic.

Endemic species, previously estimated as comprising 15-16 % of the Mediterranean fauna, are today thought to represent 28-29 % of the total in littoral waters, but the percentage decreases rapidly with depth to 15 % at 200 m, 13 % at 500 m, 14 % at 1,000 m and 19 % at 2,000 m (BELLAN-SANTINI *et alii*, 1992). Thus, with increasing depth the originality of the fauna decreases. It is difficult to propose a valid hypothesis to explain the origin of this endemism, mainly because more than 75 % of endemic deep-sea species are known only by a very small number of individuals recorded in one or two stations. Often the deep water endemic species are very similar to species in the northeastern Atlantic, thus posing questions about their taxonomic relationships and their true geographical distribution. With few exceptions, at the genus level endemic genera are almost exclusively littoral. Accepting that a parallelism exists between the hierarchical rank of an endemic taxon and the time this taxon has been in its new environment allows this conclusion: the almost total absence of endemic genera and families in the bathyal and abyssal faunas clearly demonstrates their youth.

In addition, the deep-water homothermy and elevated salinity (more than 38 psu) constitute a barrier for a large number of species. With the addition of the brutal Quaternary climatic fluctuations, the relative poverty and the low degree of originality in the Mediterranean fauna can easily be understood. Cold stenothermal species are not represented and several typically deep-sea taxonomic groups, *e.g.* hexactinellid sponges and holothurid elasipods, are very rare in the Mediterranean Sea. The main stocks of species are shown in figure 4.

## Geographical changes

The origin and the importance of Mediterranean species (no data specifically for deep-water species) are the following:

| | |
|---|---|
| Atlantic origin | 50.2 % |
| Endemic species | 28.6 % |
| Atlantico-Pacific origin | 16.8 % |
| Indo-Pacific origin | 4.4 % |

The Mediterranean deep-water fauna consists mainly of eurybathic species (Fig. 4) with an extensive geographical distribution that is in accordance with this postulate: the deeper their Mediterranean bathymetric location, the more far-reaching their geographical range elsewhere. Thus, about 70 % of these species (only 20 % below 1,000 m) occur also in the boreal province of the Atlantic Ocean where they are generally neritic, *e.g.* the echinoderms *Leptometra celtica* and *Plutonaster bifrons*, the crustacean *Amphilepsis norvegica*, the polychaete *Nephtys ciliata*, the molluscs *Dentalium agile*, *Anamathia rissoana*, *etc*.). About 35 % of the species are known along both the East and West coasts of the North Atlantic Ocean. The close affinity between Mediterranean and Atlantic congeneric deep-water species suggests that the ancestors of the Mediterranean bathyal endemic species came when conditions were favourable from the Atlantic where they lived in a similar habitat. This demonstrates that the Gibraltar sill has not always been an impenetrable barrier to colonisation or recolonisation of the deeper waters of the Mediterranean, for strictly deep-dwelling species, *e.g.* the decapod crustacean family Polychelida, a small exclusively deep-water group, considered as the sole "living fossil" in the Mediterranean (DURISH, 1987; ABELLÓ and CARTES, 1992). However, typically bathyal or abyssal taxonomical groups are absent, as well as cold water stenothermal species that elsewhere represent the major part of the deep-sea fauna.

For the Mediterranean fauna as a whole, the general tendency is an impoverishment in species from West to East (Table 3). Species occurring in the both western and eastern basins are always encountered more deeply in the eastern basin. This appears to be a general tendency in all oceans, the bathymetric distribution is always deeper on the east coast than on the west coast (ZEZINA, 1987).

The greater qualitative impoverishment of the Mediterranean deep-sea fauna can be explained in several ways:

- The main direction for colonization is West to East, It encounters successive difficulties, *i.e.* the passage of the Gibraltar and of the Siculo-Tunisian sills, and also the transit of the western basin where conditions differ slightly from those of the eastern basin, thus, preventing, in particular, the establishment of of a subtropical fauna (TAVIANI, 2002).

- The validity of factual interpretation is linked to the volume of data collected. It is less in the eastern basin.

- The successive climatic changes during the Quaternary and their consequent faunal replacements may be a partial explanation for the increase in impoverishment eastward along the route of colonization.

- The Lessepsian migration from the Red Sea through the Suez canal to the Mediterranean contributed nothing to the deep-sea fauna.

## Conclusions

The existing deep Mediterranean appears to be much younger than any other of the world's deep oceans. Its fauna is composed mainly of primitive taxonomic groups among the phyla represented, whereas a small fraction of specialized taxa exists in the deep-sea fauna.

The Mediterranean is an important centre of evolution with future speciations more than likely, particularly of new endemic species, because of the strong gradients among the diversity of its environmental particularities. The Mediterranean appears to be a remarkable natural laboratory for the study of the processes of recent colonization as related to the unique history of each of the two great Mediterranean basins.

## Acknowledgments

We are indebted to Nestor J. SANDER (USA) for his comments and suggestions regarding the presentation in English, as well as to the three referees Jean-Claude SORBE (Station Marine d'Arcachon, Université de Bordeaux 1, France), Alain COUTELLE (Département des Sciences de la Terre, Université de Bretagne occidentale, Brest, France) and Jean-Loup RUBINO (Centre Scientifique et Technique, Total SA, Pau) for their helpful suggestions and constructive comments.